\documentclass[11pt]{article}

\usepackage[utf8]{inputenc}
\usepackage[none]{hyphenat} % añadido para evitar que LaTeX corte las palabras
\usepackage{amsmath}
\usepackage{amssymb}
\usepackage{amsfonts}
\usepackage{multirow}
\usepackage{natbib}
\usepackage{enumerate}
\usepackage{hyperref}
\usepackage{graphicx}
\hypersetup{colorlinks=true, linkcolor=blue, urlcolor=blue, citecolor=blue}
\usepackage{listings}
\usepackage{color}

\usepackage[top=2cm, bottom=2cm, left=2.5cm, right=2.5cm]{geometry}

\title{Enlarging of the sample to address multicollinearity}
\author{
    \bf{Román Salmerón Gómez}\thanks{Professor, Department of Quantitative methods for economics and business, University of Granada, Spain (e-mail: romansg@ugr.es).}
    \and
    \bf{Catalina García García}\thanks{Professor, Department of Quantitative methods for economics and business, University of Granada, Spain (e-mail: cbgarcia@ugr.es).}
    \and 
    \bf{Ainara Rodríguez Sánchez}\thanks{Professor, Department of Applied Economics and Economic History, National University of Distance Education, Madrid, Spain  (e-mail: arsanchez@cee.uned.es).}}
   \date{\today}

\title{Unraveling Residualization: enhancing its application and exposing its relationship with the FWL theorem}

\author{
    \bf{Catalina Garc\'ia Garc\'ia}\thanks{Department of Quantitative methods for economics and business, University of Granada, Spain (e-mail: cbgarcia@ugr.es).}
    \and
    \bf{Rom\'an Salmer\'on G\'omez}\thanks{Department of Quantitative methods for economics and business, University of Granada, Spain (e-mail: romansg@ugr.es).}
    \and 
    \bf{Claudia Garc\'ia Garc\'ia}\thanks{Department of Applied Economics, Structure and History, Complutense University of Madrid (e-mail: clgarc13@ucm.es).}
}
\date{\today}

\begin{document}

%%%%%%%%%%%%%%%%%%%%%%%%%%%%%%%%%%%%%%%%%%%%%%%%%%%%%

%\begin{frontmatter}

\maketitle

% Abstract and keywords 
\begin{abstract}
The residualization procedure has been applied in many different fields to estimate models with multicollinearity. However, there exists a lack of understanding of this methodology and some authors discourage its use. This paper aims to contribute to a better understanding of the residualization procedure to promote an adequate application and interpretation of it among statistics and data sciences. We highlight its interesting potential application, not only to mitigate multicollinearity but also when the study is oriented to the analysis of the isolated effect of independent variables. The relation between the residualization methodology and the Frisch-Waugh-Lovell (FWL) theorem is also analyzed, concluding that, although both provide the same estimations, the interpretation of the estimated coefficients is different. These different interpretations justify the application of the residualization methodology regardless of the FWL theorem. A real data example is presented for a better illustration of the contribution of this paper. \\
keywords: \textit{residualization, multicollinearity, FWL theorem, isolated effect}
\end{abstract}

%\begin{keyword}
%Residualization \sep Data sciences \sep Multicollinearity \sep FWL theorem \sep Isolated effect
%%\JEL C01 \sep C13 \sep C29 \sep C59
%\end{keyword}

%\end{frontmatter}

%% \linenumbers

%%%%%%%%%%%%%%%%%%%%%%%%%%%%%%%%%%%%%%%%%%%%%%%%%%%%%
%% main text

\section{Introduction}

The Frisch-Waugh-Lovell (FWL) theorem, first published by Ragnar Frisch and Frederick V. Waugh in 1933 (\cite{FrischWaugh1933}) and further demonstrated by Michael Lovell in 1963 (\cite{Lovell1963,Lovell2008}), is a theorem that allows the researcher to reduce multivariate regressions to univariate ones. This is extremely useful when the relationship between some variables is concerned in an empirical study, but other factors must be controlled. This method shows that there are two additional ways for obtaining the estimation of a particular parameter of a regression model, both based on some modifications of the variables included in the model. 

At this point, the residualization procedure emerges. Residualization appears, mainly, to mitigate the potential multicollinearity of a regression model, and it can be interpreted as a similar procedure to the FWL theorem. The method is briefly explained by \cite{hill2007collinearity}, who based their research on works by \cite{kennedy1982eliminating} and \cite{buse1994brickmaking}, and was fully developed recently by \cite{Garciaetal2019b}, who present the algebraical expressions for its estimation, individual and joint inference, the goodness of fit and the effect that its applications have on the degree of multicollinearity existing in the model, as well as the advantages of using it instead of methodologies such as ridge regression or principal components analysis, the use of which, although widespread (e.g. \cite{Adleretal2010,Bulut2020,Menonetal2023,Poldaru2014,Rao2023forecasting,Vinci2022,Weietal2023decoupling}) has some drawbacks (see \cite{GarciaTesis2020}). This methodology has been applied in numerous studies published in relevant social science journals in many different fields, such as linguistics, environmental issues and economic development and policies; see, for example, \cite{Ambridgeetal2012,BandeljMahutga2010,Bradshaw1987,CohenGoldberg2012,DaSilva2022,Jaeger2010,Jorgenson2006,jorgenson2014economic,JorgensonBurns2007,JorgensonClark2009,jorgenson2018inequality,jorgenson2015changing,KentorKick2008,knight2011environmental,Kupermanetal2008,Kupermanetal2010,Lemhoferetal2008,MahutgaBandelj2008,rommers2015verbal,WaltonRagin1990}.

Regarding the multicollinearity issue, it is important to note that it may arise in two ways: perfect or exact multicollinearity (it is not possible to obtain a unique solution for the OLS estimations) and imperfect or near multicollinearity (it is possible to obtain a unique solution for the OLS estimations but other problems emerge). With near multicollinearity, the OLS estimators may present relevant difficulties (see \cite{GarciaTesis2020} for more information), such as: (1) high variances of the estimators; (2) tendency to consider the estimated parameters as non-significant; (3) greater confidence intervals; (4) non-robust results: the estimates have a high sensitivity to small changes in the initial data; (5) a considerable possibility of the appearance of incorrect signs for the estimated coefficients, and (6) difficulty in fixing the individual effects of the independent variables to the explained variable, and hence, to the explained sum of squares and the coefficient of determination, that is, difficulty in isolating the effect of the explanatory variables on the response variable. This paper focuses on this last challenge.

Both methodologies (FWL theorem and residualization) are connected since one of the ways presented for obtaining the estimated parameter with the FWL theorem partially coincides with the residualization procedure. However, it will be shown that the aims of both procedures are different: while the FWL theorem appears for understanding multiple and partial regression and the obtention of the estimated coefficients in different ways, the residualization procedure focuses its attention on the appearance of strong multicollinearity problems and their mitigation, with some novelties on the interpretation of the obtained coefficients that are different than those from the FWL theorem. 

In sum, if the FWL theorem shows that there exist different ways to obtain the same estimated coefficient, the residualization transforms the original model into another model that is practically similar but with differences that make it very interesting. On the one hand, residualization mitigates multicollinearity and, on the other hand, it provides a new interpretation for the estimated model. This paper aims to contribute to a better understanding of the residualization procedure to promote an adequate application and interpretation of it.

The structure of the paper is as follows: Section \ref{new_interpretation} presents the residualization methodology and shows the potential application of residualization to correctly quantify the effect of the explanatory variable on the explained variable when a worrisome multicollinearity exists. The relationship of this procedure with the FWL theorem is also addressed. Section \ref{example} illustrates the procedure and implications with a real data example and, finally, Section \ref{conclusion} summarizes the main conclusions.

%%%%%%%%%%%%%%%%%%%%%%%

\section{Residualization and the FWL theorem: similarities and differences}
    \label{new_interpretation}

First, the residualization method is going to be explained. To that end, the following multiple linear regression model is specified ($\mathbf{X}_{1}=\mathbf{1}$, with $\mathbf{1}$ being a vector of ones with appropriate dimensions):
    \begin{eqnarray}
        \mathbf{y} &=& \beta_{1} \mathbf{X}_{1} + \beta_{2} \mathbf{X}_{2} + \dots + \beta_{i} \mathbf{X}_{i} + \dots + \beta_{p} \mathbf{X}_{p} + \mathbf{u} \nonumber\\ 
             &=& \beta_{1} + \beta_{2} \mathbf{X}_{2} + \dots + \beta_{i} \mathbf{X}_{i} + \dots + \beta_{p} \mathbf{X}_{p} + \mathbf{u},\label{modelo1}
    \end{eqnarray}
where $\mathbf{u}$ is a random disturbance, $p$ is the number of explanatory variables, and we have $n$ observations. 

Suppose, to simplify the procedure, that we are working with model (\ref{modelo1}) for $p=3$:
    \begin{equation}
        \mathbf{y} = \beta_{1} + \beta_{2} \mathbf{X}_{2} + \beta_{3} \mathbf{X}_{3} + \mathbf{u}, \label{modelo.p3}
    \end{equation}
    and we are interested in analyzing the influence of $\mathbf{X}_{3}$ on $\mathbf{y}$. If variables $\mathbf{X}_{2}$ and $\mathbf{X}_{3}$ are linearly related, when one of them increases/diminishes, the other will also change (in the same or the opposite direction). With this situation, the condition \textit{ceteris paribus} will not be verified\footnote{See the Appendix for more information.}, and hence:
    $$\frac{\partial \mathbf{y}}{\partial \mathbf{X}_{3}} = \beta_{3} + \beta_{2} \frac{\partial \mathbf{X}_{2}}{\partial \mathbf{X}_{3}}.$$
    
    In the same line, if we are interested in analyzing the influence of $\mathbf{X}_{2}$ on $\mathbf{y}$, and $\mathbf{X}_{2}$ and $\mathbf{X}_{3}$ are linearly related, then:
    $$\frac{\partial \mathbf{y}}{\partial \mathbf{X}_{2}} = \beta_{2} + \beta_{3} \frac{\partial \mathbf{X}_{3}}{\partial \mathbf{X}_{2}}.$$
    
    Therefore, we cannot separate the individual influence of each explanatory variable and the results obtained from the ordinary least squares (OLS) estimation for model (\ref{modelo.p3}) will not reflect the actual effect of each one - the variables are not independent from each other. This fact implies the existence of multicollinearity\footnote{There are particular indicators that reflect the existence of strong collinearity problems, such as the Variance Inflation Factor (VIF). See the Appendix for more information.}. 
    
    To solve the ``\textit{ceteris paribus} problem'' (i.e. mitigating multicollinearity), let us specify the following auxiliary model:
    \begin{equation}
    \mathbf{X}_2=\boldsymbol{\alpha}_1+\boldsymbol{\alpha}_2\mathbf{X}_{3}+\mathbf{v}.    \label{reg.auxiliar1}
\end{equation}
    
    The auxiliary regression (\ref{reg.auxiliar1}) shows the linear relationship between $\mathbf{X}_{2}$ and $\mathbf{X}_{3}$, and from the OLS estimation of model (\ref{reg.auxiliar1}), we will obtain that variable $\mathbf{X}_{2}$ can be expressed as $\widehat{\alpha}_{1} + \widehat{\alpha}_{2} \mathbf{X}_{3} + \mathbf{e}_{2}$, where $\mathbf{e}_{2}$ represent the estimated residuals.
    
    Hence, the initial model (\ref{modelo.p3}) can be retyped as:
    \begin{eqnarray}
        \mathbf{y} &=& \beta_{1} + \beta_{2} \mathbf{X}_{2} + \beta_{3} \mathbf{X}_{3} + \mathbf{u} = \beta_{1} + \beta_{2} (\widehat{\alpha}_{1} + \widehat{\alpha}_{2} \mathbf{X}_{3} + \mathbf{e}_{2})  + \beta_{3} \mathbf{X}_{3} + \mathbf{u} \nonumber \\
            &=& (\beta_{1} + \beta_{2} \widehat{\alpha}_{1}) + \beta_{2} \mathbf{e}_{2} + (\beta_{3} + \beta_{2} \widehat{\alpha}_{2} ) \mathbf{X}_{3}  + \mathbf{u}. \label{reparametrizacion}
    \end{eqnarray}
  
 The previous expression is known as the the residualized model, which can be expressed as:
  \begin{equation}
     \mathbf{y} = \delta_{1} + \delta_{2} \mathbf{e}_{2} + \delta_{3} \mathbf{X}_{3} + \mathbf{u}. \label{modelo.residualized}
\end{equation}
By comparing expression (\ref{reparametrizacion}) with (\ref{modelo.residualized}), it can be observed that $\delta_{1}=\beta_{1} + \beta_{2} \widehat{\alpha}_{1}$, $\delta_{2}=\beta_{2}$ and $\delta_{3}=\beta_{3} + \beta_{2} \widehat{\alpha}_{2}$. Thus, the estimation of the coefficient of the residualized variable (i.e. the explained variable in the auxiliary regression (\ref{reg.auxiliar1}), $\mathbf{X}_{2}$) is not altered ($\widehat{\delta}_{2} = \widehat{\beta}_{2}$), as \cite{Garciaetal2019b} demonstrated for $p$ explanatory variables.

       The corresponding residuals from (\ref{reg.auxiliar1}), $\mathbf{e}_{2}$, will represent the part of the variable $\mathbf{X}_2$ that has no relation with the other exogenous variable of the original model (\ref{modelo.p3}) since the residuals are orthogonal to $\mathbf{X}_{3}$, that is, $\mathbf{e}_{2}^{t} \mathbf{X}_{3} = \mathbf{0}$, with $\mathbf{0}$ being a vector of zeros with appropriate dimensions. Then, in this case, as $\mathbf{e}_{2}$ and $\mathbf{X}_{3}$ are not linearly related, the \textit{ceteris paribus} condition is verified, and:
    $$\frac{\partial \mathbf{y}}{\partial \mathbf{e}_{2}}= \delta_{2},\ \ \ \ \ \ \ \ \frac{\partial \mathbf{y}}{\partial \mathbf{X}_{3}} = \delta_{3} + \delta_{2} \widehat{\alpha}_{2}.$$
    
    Additionally, the reader may note that the residualized model quantifies the influence of the reparametrized explanatory variables on $\mathbf{y}$, taking into account the linear relation existing between them. Thus, taking into account the linear relation between the explanatory variables involved in the auxiliary regression (\ref{reg.auxiliar1}), it is now possible to quantify the effects of both variables independently, solving the ``\textit{ceteris paribus} problem''\footnote{Note that if variables $\mathbf{X}_{2}$ and $\mathbf{X}_{3}$ are linearly independent, then $\widehat{\alpha}_{2} = 0$.}. 
    
In conclusion regarding the residualization procedure, it has been demonstrated its capability to mitigate the collinearity problem and also to isolate the marginal effects of the explanatory variables on the response variable.

On the other hand, the Frisch-Waugh-Lovell (FWL) theorem (\cite{FrischWaugh1933,Lovell1963,Lovell2008}) establishes that it is possible to obtain the same estimations from two different methods. The first method is called the ``individual trend method'' (where the tendency of the independent variables has been eliminated), and the second method is called the ``partial time regression method'' (where the tendency is not eliminated but is included as an additional independent variable). In this way, they disproved the belief that if we worked with variables with a trend, the associated coefficients would correctly measure the relationship with the dependent variable, while the coefficients of the variables in which the trend had been eliminated would not. Those same estimates are interpreted in the same way.

Methodologically, starting from the original simplified model (\ref{modelo.p3}), the FWL theorem is based on the following expression:
    \begin{equation}
        \mathbf{e}_{y} = \gamma_{1} + \gamma_{2} \mathbf{e}_{2} + \boldsymbol{\nu}, \label{modelo.FWL}
    \end{equation}
    where $\mathbf{e}_{y}$ are the residuals obtained from the OLS estimation of regression $\mathbf{y} = \theta_{1} + \theta_{2} \mathbf{X}_{3} + \mathbf{w}$, with $\gamma_{2}$ coinciding with the estimation by OLS of the coefficient $\beta_{2}$ of model (\ref{modelo.p3}), and hence, coinciding also with $\delta_{2}$ from model (\ref{reparametrizacion}). That is, $\widehat{\beta}_{2} = \widehat{\delta}_{2} = \widehat{\gamma}_{2}$. So, it can be observed that we can obtain the particular estimated parameter of a regression model by using different ways, as the FWL theorem tries to conclude.
    
    With this, we can see that both methodologies have some similarities. Note that to establish the model (\ref{modelo.FWL}), $\mathbf{e}_{2}$ should be previously obtained, as in the residualization. Then, it is clear that residualization and the FWL theorem share a methodology but lead to two different final models, (\ref{modelo.residualized}) and (\ref{modelo.FWL}), with different aims.

In fact, residualization provides an alternative interpretation for the estimated parameters, that is not taken into consideration in the literature (\cite{GarciaTesis2020}). Thus, it could be interesting to apply a residualization procedure not only to mitigate collinearity or to quantify the marginal effect of the explanatory variables, but also to obtain an alternative interpretation of the residualized variable. 

In the residualized model, the coefficient of the $j$ residualized variable is interpreted as the effect of the part of the variable not related to the rest of the explanatory variables. This differentiating fact is the main justification to apply the residualization procedure but, at the same time, is its main limitation. Thus, due to this interpretation, it is not always possible to apply the procedure. Note that this opportunity to obtain new interpretations is both an advantage but also a limitation (\cite{GarciaTesis2020}).

In short, the FWL theorem establishes equivalences between the estimates of two different models, while residualization establishes the differences between two very similar models that are not exactly the same.

%%%%%%%%%%%%%%%%%%%%%%%

\section{Real data example}
    \label{example}

\cite{NelsonSiegel1987} proposed the forward rate curve to fit the term structure using the following flexible function:
\begin{equation}
    y(\tau)=\beta_1+\beta_2\left(\frac{1-e^{-\lambda\tau}}{\lambda\tau}\right)+\beta_3\left(\frac{1-e^{-\lambda\tau}}{\lambda\tau}-e^{-\lambda\tau}\right),
    \label{model.NS}
\end{equation}
where $y(\tau)$ is the zero rate for maturity $\tau$, and $\lambda$ is a hump term that determines the location of the maximum or minimum curvature component. The parameters have an economic interpretation in the term structure. Thus, the contribution of the long-term component is $\beta_1$, that of the short-term component is $\beta_2$, and $\beta_3$ indicates the contribution of the medium-term component. The existence of collinearity in the Nelson-Siegel (NS) model is provoked by the correlation between the short- and medium-term components, and could make it difficult to obtain good estimates of the parameters, \cite{DieboldLi2006}. Note that the potential collinearity depends on the value of $\lambda$. Recently, some authors have applied ridge regression to estimate this model, e.g. \cite{Annaertetal2013} and \cite{Leonetal2018}.

The NS model is widely used by central banks and other market participants as a model for the term structure of the interest rate due to its ability to estimate yields for all maturities, the intuitive interpretations of the factor obtained and its good performance in out-of-sample forecasting. Several extensions of this model have been proposed. \cite{Svensson1994} added a second hump term, and \cite{DieboldLi2006} presented a dynamic version of the model. Since the purpose of this paper is the clarification of the residualization procedure, we will apply the original static version of \cite{NelsonSiegel1987}, although it could be interesting to extend this analysis to alternative versions of the original NS model.

For the empirical application, we use the database previously used by \cite{DieboldLi2006}, which contains the end-of-month price quotes (bid-ask average) for U.S. treasuries from January 1985 through December 2000, taken from the Center for Research in Security Prices (CRSP) government bond files. We have decided to opt for less recent but widely used data from previous works in order to make the results as robust as possible. 

Maturities are fixed at month $\tau=3, 6, 9, 12, 15,18, 21, 24, 30, 36, 48, 60, 72,$ $84, 96, 108, 120$. Additionally, since we estimate the static model, we select four dates: January 31st 1985, 1990, 1995 and 2000. \cite{DieboldLi2006} fixed the value of $\lambda$ at 0.0609, which indicates only a weak positive correlation between the factor loadings. To show the consequences of high correlations, \cite{Leonetal2018} replicated the results of \cite{DieboldLi2006} with $\lambda$ equal to 0.01. In this paper, $\lambda$ has been fixed at 0.01, which leads to a coefficient of correlation equal to $-0.9920$ between the short and medium terms.

Table \ref{olstable} shows the estimation by OLS. Note that the negative values of the long term ($\widehat{\beta}_1$) do not make economic sense, and that the intercept is not individually significant in all cases. These are common symptoms of the existence of collinearity, which is diagnosed with a VIF equal to 64.61. 

Regardless of the existence of collinearity, it may be interesting to know the isolated effect of the short and medium term. It is evident that the short term and the medium term are important variables to explain the variability of the dataset, although the medium term can be interpreted as slightly more important than the short term. In fact, it could be interpreted that the medium term includes in some way the short term. For this reason, the medium term is the variable selected to be residualized.

When the residualization is applied, the medium term is substituted by the residuals obtained from the auxiliary regression between the medium and the short terms. These residuals can be interpreted as part of the medium term that is not explained by the short term. Thus, Table \ref{orttable} presents the estimations of the residualizations. Note that the values of the long term are always positive, while the short term becomes negative. This makes economic sense. In addition, all the coefficients of the variables are individually significant with a level of confidence of 95\%, and the coefficient of determination of the original model is kept. Note that the estimations for the medium term are maintained, as the FWL theorem aims.

\begin{table}
\begin{center}
\begin{tabular}{l|l|l|l|l}

& 1985/01/31 & 1990/01/31 & 1995/01/31 & 2000/01/31 \\
\hline

\multirow{2}{*}{$\hat{\beta}_1$} & -6.29014*** & 6.2755*** & -3.4472 & -2.8649 \\
& (2.0218) & (0.8092) & (2.5332) & (2.0282)\\
\multirow{2}{*}{$\hat{\beta}_2$} & 14.3533*** & 1.6784* & 9.6265*** & 8.6828*** \\
& (1.9655) & (0.7860) & (2.4627) & (2.0241)\\
\multirow{2}{*}{$\hat{\beta}_3$} & 30.9317** & 3.6328*** & 18.5377*** & 15.1018*** \\
& (2.8875) & (1.1557) & (3.6180) & (2.9730) \\
\hline
$R^2$ & 0.9834 & 0.8384 & 0.8919 & 0.8160 \\
\hline
\end{tabular}
\end{center}
\caption{OLS estimation for different data sets (*** indicates 99\% confidence level, ** indicates 95\% confidence level and * indicates 90\% confidence level.)} \label{olstable}
\end{table}

\begin{table}
\begin{center}
{\begin{tabular}{l|l|l|l|l}

& 1985/01/31 & 1990/01/31 & 1995/01/31 & 2000/01/31 \\
\hline
\multirow{2}{*}{$\hat{\beta}_1$} & 15.2579*** & 8.8063*** & 9.4667*** & 7.6554*** \\
& (0.2033) & (0.0813) & (0.2548) & (0.2694) \\
\multirow{2}{*}{$\hat{\beta}_2$} & -6.5381*** & -0.7752*** & -2.8939*** & -1.5170*** \\
& (0.2445) & (0.0978) & (0.3063) & (0.2518)\\
\multirow{2}{*}{$\hat{\beta}_3^0$} & 30.9317** & 3.6328*** & 18.5377*** & 15.1018*** \\
& (2.8875) & (1.1557) & (3.6180) & (2.9730)\\
\hline
$R^2$ & 0.9834 & 0.8383 & 0.8918 & 0.8159 \\
\hline
\end{tabular}}
\end{center}
\caption{Residualization for different datasets (*** indicates 99\% confidence level, ** indicates 95\% confidence level and * indicates 90\% confidence level.)} \label{orttable}
\end{table}

Thus, this example shows that the residualization allows, on the one hand, an isolation of the short-term and medium-term effects on the dependent variable, since by residualizing the second one, the \textit{ceteris paribus} is verified in the interpretation of the estimation of $\beta_{2}$ and $\beta_{3}$ in (\ref{model.NS}). On the other hand, a new interpretation of $\widehat{\beta}_{3}$ is obtained that is not possible with the initial model and, therefore, allows us to answer questions different from the initial ones, and different from the FWL theorem, as has been commented in the previous section.

At the same time, by applying residualization, the estimate of $\beta_{2}$ is modified, obtaining results that are consistent with the economic theory that supports the Nelson-Siegel model, contrary to what happens with the original estimations. 

%%%%%%%%%%%%%%%%%%%%%%%

\section{Conclusions}
\label{conclusion}

Some authors, such as \cite{York}, discourage the use of residualization because it \textit{leads to biased coefficients and standard error estimates, and does not address the fundamental problem of collineari\-ty}. However, this paper shows that residualization can mitigate collinearity by allowing an analysis of the isolated effect of the individual variables.

It is also important to note that this methodology changes the interpretation of the residualized variable. Therefore, residualization should be applied only if the modified variable is interpretable.

Definitively, why should a researcher use this method? A first answer could be to mitigate the degree of multicollinearity existing in a model for a better analysis of it (for example, to isolate the effect of the residualized variable); but even if the multicollinearity were not worrisome, the researcher could be interested in applying residualization to obtain an alternative interpretation of the parameter associated with the residualized variable.

Finally, the relationship between the residualization and the FWL theorem is also addressed. It is shown that although the first can be considered part of the second, they are methodologies that have different aims and, therefore, provide different results. The most important difference is that with the residualization, the interpretation of the estimates obtained is modified, while with the FWL theorem, it is not.

%%%%%%%%%%%%%%%%%%%%%%%%%%%%%%%%%%%%%%%%%%%%%%%%%%%%%

%% Funding

\section*{Acknowledgements}
This work has been supported by project A-SEJ-496-UGR20 of the Andalusian Government's Counseling of Economic Transformation, Industry, Knowledge and Universities (Spain).

%% Appendix

%\newpage
\appendix

\section{Multicollinearity: marginal effects and detection}

\textit{Marginal effects}. Starting from model (\ref{modelo1}), the effect of the variable $\mathbf{X}_{i}$ on $\mathbf{y}$ is obtained if the rest of the explanatory variables remain constant (\textit{ceteris paribus}), as follows:
    \begin{equation}
        \label{em_cp}
        \frac{\partial \mathbf{y}}{\partial \mathbf{X}_{i}} = \beta_{i}, \quad i=2,\dots,p.
    \end{equation}
    However, under a high degree of (near) multicollinearity, the \textit{ceteris paribus} condition is not verified, and consequently, the expression (\ref{em_cp}) is not verified.

   Indeed, if $\mathbf{X}_{i}$ is linearly related to the rest of the explanatory variables (or at least one of them), it is verified that:
    \begin{equation}
        \label{em_scp}
        \frac{\partial \mathbf{y}}{\partial \mathbf{X}_{i}} = \beta_{i} + \sum \limits_{h=2,j\not=i}^{p} \beta_{h} \frac{\partial \mathbf{X}_{h}}{\partial \mathbf{X}_{i}}, \quad i=2,\dots,p.
    \end{equation}
    By comparing expressions (\ref{em_cp}) and (\ref{em_scp}), it is observed that the existence of a worrisome multicollinearity impedes the correct quantification of the variable $\mathbf{X}_{i}$ on $\mathbf{y}$, i.e. its marginal effect cannot be correctly quantified.
    
\textit{Detection}. The variance inflation factor (VIF) is one of the most applied techniques to detect strong (near) multicollinearity problems in a model, and can be defined as the percentage of the variance that is inflated for each coefficient:
\begin{equation}\label{vifi}
\text{VIF}_j = \frac{\widehat{Var}(\widehat{\beta}_j)}{\widehat{Var}(\widehat{\beta}_{j \mathbf{O}})} = \frac{1}{1-R^2_j}, \quad j=2,\dots,p,
\end{equation}
where $R^2_j$ is the coefficient of determination from regression $\mathbf{X}_j=\mathbf{X}_{-j}\boldsymbol{\alpha}+\mathbf{v}$.
The subscript $\mathbf{O}$ represents the orthogonal (no collinearity) situation. \cite{Marquardt1970} stated that a value lower than 10 indicates that there is no problematic collinearity, although some authors set the VIF threshold to 4 (\cite{OBrien2007}).

%\newpage
%% References
\bibliographystyle{chicago}
\bibliography{bib}

\begin{thebibliography}{}

\bibitem[\protect\citeauthoryear{Adler, Yazhemsky, and Tarverdyan}{Adler
  et~al.}{2010}]{Adleretal2010}
Adler, N., E.~Yazhemsky, and R.~Tarverdyan (2010).
\newblock A framework to measure the relative socio-economic performance of
  developing countries.
\newblock {\em Socio-Economic Planning Sciences\/}~{\em 44\/}(2), 73--88.

\bibitem[\protect\citeauthoryear{Ambridge, Pine, and Rowland}{Ambridge
  et~al.}{2012}]{Ambridgeetal2012}
Ambridge, B., J.~Pine, and C.~Rowland (2012).
\newblock {Semantics versus statistics in the retreat from locative
  overgeneralization errors}.
\newblock {\em Cognition\/}~{\em 123\/}(2), 260--279.

\bibitem[\protect\citeauthoryear{Annaert, Claes, De~Ceuster, and Zhang}{Annaert
  et~al.}{2013}]{Annaertetal2013}
Annaert, J., A.~Claes, M.~De~Ceuster, and H.~Zhang (2013).
\newblock {Estimating the spot rate curve using the Nelson--Siegel model: A
  ridge regression approach}.
\newblock {\em International Review of Economics \& Finance\/}~{\em 27},
  482--496.

\bibitem[\protect\citeauthoryear{Bandelj and Mahutga}{Bandelj and
  Mahutga}{2010}]{BandeljMahutga2010}
Bandelj, N. and M.~Mahutga (2010).
\newblock {How socio-economic change shapes income inequality in post-socialist
  Europe}.
\newblock {\em Social Forces\/}~{\em 88\/}(5), 2133--2161.

\bibitem[\protect\citeauthoryear{Bradshaw}{Bradshaw}{1987}]{Bradshaw1987}
Bradshaw, Y. (1987).
\newblock {Urbanization and underdevelopment: A global study of modernization,
  urban bias, and economic dependency}.
\newblock {\em American Sociological Review\/}~{\em 52\/}(2), 224--239.

\bibitem[\protect\citeauthoryear{Bulut}{Bulut}{2020}]{Bulut2020}
Bulut, H. (2020).
\newblock The construction of a composite index for general satisfaction in
  turkey and the investigation of its determinants.
\newblock {\em Socio-Economic Planning Sciences\/}~{\em 71}, 100811.

\bibitem[\protect\citeauthoryear{Buse}{Buse}{1994}]{buse1994brickmaking}
Buse, A. (1994).
\newblock Brickmaking and the collinear arts: A cautionary tale.
\newblock {\em Canadian Journal of Economics\/}, 408--414.

\bibitem[\protect\citeauthoryear{Cohen-Goldberg}{Cohen-Goldberg}{2012}]{CohenGoldberg2012}
Cohen-Goldberg, A. (2012).
\newblock {Phonological competition within the word: Evidence from the phoneme
  similarity effect in spoken production}.
\newblock {\em Journal of Memory and Language\/}~{\em 67\/}(1), 184--198.

\bibitem[\protect\citeauthoryear{Da~Silva, Costa, and Lopes-Ahn}{Da~Silva
  et~al.}{2022}]{DaSilva2022}
Da~Silva, A.~V., M.~A. Costa, and A.~L. Lopes-Ahn (2022).
\newblock Accounting multiple environmental variables in dea energy
  transmission benchmarking modelling: The 2019 brazilian case.
\newblock {\em Socio-Economic Planning Sciences\/}~{\em 80}, 101162.

\bibitem[\protect\citeauthoryear{Diebold and Li}{Diebold and
  Li}{2006}]{DieboldLi2006}
Diebold, F. and C.~Li (2006).
\newblock {Forecasting the term structure of government bond yields}.
\newblock {\em Journal of econometrics\/}~{\em 130\/}(2), 337--364.

\bibitem[\protect\citeauthoryear{Frisch and Waugh}{Frisch and
  Waugh}{1933}]{FrischWaugh1933}
Frisch, R. and F.~V. Waugh (1933).
\newblock Partial time regressions as compared with infividual trends.
\newblock {\em Econometrica\/}~{\em 1\/}(4), 387--401.

\bibitem[\protect\citeauthoryear{Garc{\'\i}a, Salmer{\'o}n, Garc{\'\i}a, and
  Garc\'{i}a}{Garc{\'\i}a et~al.}{2019}]{Garciaetal2019b}
Garc{\'\i}a, C.~B., R.~Salmer{\'o}n, C.~Garc{\'\i}a, and J.~Garc\'{i}a (2019).
\newblock {Residualization: justification, properties and application}.
\newblock {\em Journal of Applied Statistics\/}.

\bibitem[\protect\citeauthoryear{Garc\'ia-Garc\'ia}{Garc\'ia-Garc\'ia}{2020}]{GarciaTesis2020}
Garc\'ia-Garc\'ia, C. (2020).
\newblock {\em {Generalization of the residualization procedure. Properties and
  environmental applications}}.
\newblock Ph.\ D. thesis, {University of Granada}.

\bibitem[\protect\citeauthoryear{Hill and Adkins}{Hill and
  Adkins}{2007}]{hill2007collinearity}
Hill, R. and L.~Adkins (2007).
\newblock Collinearity. in a companion to theoretical econometrics.
\newblock {\em Oxford, UK: Blackwell Publishing. Hausman, JA (1978).
  Specification Test in Econometrics. ECONOMETRICA, Journal of Econometric
  Society\/}~{\em 46\/}(6), 1251--1271.

\bibitem[\protect\citeauthoryear{Jaeger}{Jaeger}{2010}]{Jaeger2010}
Jaeger, T. (2010).
\newblock {Redundancy and reduction: Speakers manage syntactic information
  density}.
\newblock {\em Cognitive psychology\/}~{\em 61\/}(1), 23--62.

\bibitem[\protect\citeauthoryear{Jorgenson}{Jorgenson}{2006}]{Jorgenson2006}
Jorgenson, A. (2006).
\newblock {Global warming and the neglected greenhouse gas: A cross-national
  study of the social causes of methane emissions intensity, 1995}.
\newblock {\em Social Forces\/}~{\em 84\/}(3), 1779--1798.

\bibitem[\protect\citeauthoryear{Jorgenson}{Jorgenson}{2014}]{jorgenson2014economic}
Jorgenson, A. (2014).
\newblock Economic development and the carbon intensity of human well-being.
\newblock {\em Nature Climate Change\/}~{\em 4\/}(3), 186.

\bibitem[\protect\citeauthoryear{Jorgenson and Burns}{Jorgenson and
  Burns}{2007}]{JorgensonBurns2007}
Jorgenson, A. and T.~Burns (2007).
\newblock {The political-economic causes of change in the ecological footprints
  of nations, 1991--2001: a quantitative investigation}.
\newblock {\em Social Science Research\/}~{\em 36\/}(2), 834--853.

\bibitem[\protect\citeauthoryear{Jorgenson and Clark}{Jorgenson and
  Clark}{2009}]{JorgensonClark2009}
Jorgenson, A. and B.~Clark (2009).
\newblock {The economy, military, and ecologically unequal exchange
  relationships in comparative perspective: a panel study of the ecological
  footprints of nations, 1975-2000}.
\newblock {\em Social Problems\/}~{\em 56\/}(4), 621--646.

\bibitem[\protect\citeauthoryear{Jorgenson, Dietz, and Kelly}{Jorgenson
  et~al.}{2018}]{jorgenson2018inequality}
Jorgenson, A., T.~Dietz, and O.~Kelly (2018).
\newblock Inequality, poverty, and the carbon intensity of human well-being in
  the united states: a sex-specific analysis.
\newblock {\em Sustainability Science\/}~{\em 13\/}(4), 1167--1174.

\bibitem[\protect\citeauthoryear{Jorgenson and Givens}{Jorgenson and
  Givens}{2015}]{jorgenson2015changing}
Jorgenson, A. and J.~Givens (2015).
\newblock The changing effect of economic development on the consumption-based
  carbon intensity of well-being, 1990--2008.
\newblock {\em PloS one\/}~{\em 10\/}(5), e0123920.

\bibitem[\protect\citeauthoryear{Kennedy}{Kennedy}{1982}]{kennedy1982eliminating}
Kennedy, P.~E. (1982).
\newblock Eliminating problems caused by multicollinearity: A warning.
\newblock {\em The Journal of Economic Education\/}~{\em 13\/}(1), 62--64.

\bibitem[\protect\citeauthoryear{Kentor and Kick}{Kentor and
  Kick}{2008}]{KentorKick2008}
Kentor, J. and E.~Kick (2008).
\newblock {Bringing the military back in: Military expenditures and economic
  growth 1990 to 2003}.
\newblock {\em Journal of World-Systems Research\/}~{\em 14\/}(2), 142--172.

\bibitem[\protect\citeauthoryear{Knight and Rosa}{Knight and
  Rosa}{2011}]{knight2011environmental}
Knight, K.~W. and E.~A. Rosa (2011).
\newblock The environmental efficiency of well-being: A cross-national
  analysis.
\newblock {\em Social Science Research\/}~{\em 40\/}(3), 931--949.

\bibitem[\protect\citeauthoryear{Kuperman, Bertram, and Baayen}{Kuperman
  et~al.}{2008}]{Kupermanetal2008}
Kuperman, V., R.~Bertram, and R.~Baayen (2008).
\newblock {Morphological dynamics in compound processing}.
\newblock {\em Language and Cognitive Processes\/}~{\em 23\/}(7-8), 1089--1132.

\bibitem[\protect\citeauthoryear{Kuperman, Bertram, and Baayen}{Kuperman
  et~al.}{2010}]{Kupermanetal2010}
Kuperman, V., R.~Bertram, and R.~Baayen (2010).
\newblock {Processing trade-offs in the reading of Dutch derived words}.
\newblock {\em Journal of Memory and Language\/}~{\em 62\/}(2), 83--97.

\bibitem[\protect\citeauthoryear{Lemh{\"o}fer, Dijkstra, Schriefers, Baayen,
  Grainger, and Zwitserlood}{Lemh{\"o}fer et~al.}{2008}]{Lemhoferetal2008}
Lemh{\"o}fer, K., T.~Dijkstra, H.~Schriefers, R.~Baayen, J.~Grainger, and
  P.~Zwitserlood (2008).
\newblock {Native language influences on word recognition in a second language:
  A megastudy}.
\newblock {\em Journal of Experimental Psychology: Learning, Memory, and
  Cognition\/}~{\em 34\/}(1), 12.

\bibitem[\protect\citeauthoryear{Le{\'o}n, Rubia, and Sanchis-Marco}{Le{\'o}n
  et~al.}{2018}]{Leonetal2018}
Le{\'o}n, A., A.~Rubia, and L.~Sanchis-Marco (2018).
\newblock {On multicollinearity and the value of the shape parameter in the
  term structure Nelson-Siegel model}.
\newblock {\em Aestimatio\/}~{\em 16}, 8.

\bibitem[\protect\citeauthoryear{Lovell}{Lovell}{1963}]{Lovell1963}
Lovell, M.~C. (1963).
\newblock Seasonal adjustment of economic time series and multiple regression
  analysis.
\newblock {\em Journal of the American Statistical Association\/}~{\em
  58\/}(304), 993--1010.

\bibitem[\protect\citeauthoryear{Lovell}{Lovell}{2008}]{Lovell2008}
Lovell, M.~C. (2008).
\newblock A simple proof of the fwl theorem.
\newblock {\em The Journal of Economic Education\/}~{\em 39\/}(1), 88--91.

\bibitem[\protect\citeauthoryear{Mahutga and Bandelj}{Mahutga and
  Bandelj}{2008}]{MahutgaBandelj2008}
Mahutga, M. and N.~Bandelj (2008).
\newblock {Foreign investment and income inequality: The natural experiment of
  Central and Eastern Europe}.
\newblock {\em International Journal of Comparative Sociology\/}~{\em 49\/}(6),
  429--454.

\bibitem[\protect\citeauthoryear{Marquardt}{Marquardt}{1970}]{Marquardt1970}
Marquardt, D. (1970).
\newblock {Generalized inverses, ridge regression, biased linear estimation and
  nonlinear estimation}.
\newblock {\em Technometrics\/}~{\em 12\/}(3), 591--612.

\bibitem[\protect\citeauthoryear{Menon, Sahadev, Mahanty, Praveensal, and
  Asha}{Menon et~al.}{2023}]{Menonetal2023}
Menon, B.~G., S.~Sahadev, A.~Mahanty, C.~Praveensal, and G.~Asha (2023).
\newblock Trivariate causality between economic growth, energy consumption, and
  carbon emissions: empirical evidence from india.
\newblock {\em Energy Efficiency\/}~{\em 16\/}(5), 41.

\bibitem[\protect\citeauthoryear{Nelson and Siegel}{Nelson and
  Siegel}{1987}]{NelsonSiegel1987}
Nelson, C. and A.~Siegel (1987).
\newblock {Pparsimonious modeling of yield curves}.
\newblock {\em q Journal of Business\/}~{\em 60\/}(4), 473--489.

\bibitem[\protect\citeauthoryear{O'Brien}{O'Brien}{2007}]{OBrien2007}
O'Brien, R. (2007).
\newblock {A caution regarding rules of thumb for variance inflation factors}.
\newblock {\em Quality \& Quantity\/}~{\em 41}, 673--690.

\bibitem[\protect\citeauthoryear{P{\~o}ldaru and Roots}{P{\~o}ldaru and
  Roots}{2014}]{Poldaru2014}
P{\~o}ldaru, R. and J.~Roots (2014).
\newblock A pca--dea approach to measure the quality of life in estonian
  counties.
\newblock {\em Socio-Economic Planning Sciences\/}~{\em 48\/}(1), 65--73.

\bibitem[\protect\citeauthoryear{Rao, Huang, Chen, Goh, and Hu}{Rao
  et~al.}{2023}]{Rao2023forecasting}
Rao, C., Q.~Huang, L.~Chen, M.~Goh, and Z.~Hu (2023).
\newblock Forecasting the carbon emissions in hubei province under the
  background of carbon neutrality: A novel stirpat extended model with ridge
  regression and scenario analysis.
\newblock {\em Environmental Science and Pollution Research\/}~{\em 30\/}(20),
  57460--57480.

\bibitem[\protect\citeauthoryear{Rommers, Meyer, and Huettig}{Rommers
  et~al.}{2015}]{rommers2015verbal}
Rommers, J., A.~S. Meyer, and F.~Huettig (2015).
\newblock Verbal and nonverbal predictors of language-mediated anticipatory eye
  movements.
\newblock {\em Attention, Perception, \& Psychophysics\/}~{\em 77\/}(3),
  720--730.

\bibitem[\protect\citeauthoryear{Svensson}{Svensson}{1994}]{Svensson1994}
Svensson, L. (1994).
\newblock {Estimating and interpreting forward interest rates: Sweden
  1992-1994}.
\newblock Technical report, National Bureau of Economic Research.

\bibitem[\protect\citeauthoryear{Vinci, Bartolacci, Salvia, and Salvati}{Vinci
  et~al.}{2022}]{Vinci2022}
Vinci, S., F.~Bartolacci, R.~Salvia, and L.~Salvati (2022).
\newblock Housing markets, the great crisis, and metropolitan gradients:
  insights from greece, 2000--2014.
\newblock {\em Socio-Economic Planning Sciences\/}~{\em 80}, 101171.

\bibitem[\protect\citeauthoryear{Walton and Ragin}{Walton and
  Ragin}{1990}]{WaltonRagin1990}
Walton, J. and C.~Ragin (1990).
\newblock {Global and national sources of political protest: Third world
  responses to the debt crisis}.
\newblock {\em American Sociological Review\/}~{\em 55}, 876--890.

\bibitem[\protect\citeauthoryear{Wei, Wei, and Liu}{Wei
  et~al.}{2023}]{Weietal2023decoupling}
Wei, Z., K.~Wei, and J.~Liu (2023).
\newblock Decoupling relationship between carbon emissions and economic
  development and prediction of carbon emissions in henan province: based on
  tapio method and stirpat model.
\newblock {\em Environmental Science and Pollution Research\/}~{\em 30\/}(18),
  52679--52691.

\bibitem[\protect\citeauthoryear{York}{York}{2012}]{York}
York, R. (2012).
\newblock Residualization is not the answer: Rethinking how to address
  multicollinearity.
\newblock {\em Social Science Research\/}~{\em 41\/}(6), 1379--1386.

\end{thebibliography}

\end{document}